\documentclass[12pt,a4paper]{article}
\usepackage{amssymb}
\usepackage{epsf}

\begin{document}

\author{Ariel M\'{e}gevand \thanks{
\ megevand@cab.cnea.gov.ar} \\
Centro At\'{o}mico Bariloche and Instituto Balseiro, \\
Comisi\'{o}n Nacional de Energ\'{\i }a At\'{o}mica \\
and Universidad Nacional de Cuyo,\\
8400 San Carlos de Bariloche, Argentina}
\title{Development of the electroweak phase transition and baryogenesis}
\maketitle

\begin{abstract}
We investigate the evolution of the electroweak phase transition, using a
one-Higgs effective potential that can be regarded as an approximation for
the Minimal Supersymmetric Standard Model. The phase transition occurs in a
small interval around a temperature $T_{t}$ below the critical one. We
calculate this temperature as a function of the parameters of the potential
and of a damping coefficient related to the viscosity of the plasma. The
parameters that are relevant for baryogenesis, such as the velocity and
thickness of the walls of bubbles and the value of the Higgs field inside
them, change significantly in the range of temperatures where the
first-order phase transition can occur. However, we find that in the likely
interval for $T_{t}$ there is no significant variation of these parameters.
Furthermore, the temperature $T_{t}$ is in general not far below the
temperature at which bubbles begin to nucleate.
\end{abstract}

\section{Introduction}

The electroweak phase transition is an appealing scenario for the generation
of the baryon asymmetry of the universe (BAU), since it contains the three
necessary ingredients known as Sakharov's conditions, namely, baryon number
violation, $C$ and $CP$ violation, and a departure from thermal equilibrium.
However, to obtain a quantitatively successful electroweak baryogenesis, one
has to consider some extension of the Standard Model (SM) in which there is
enough $CP$ violation as well as a sufficiently strong first-order phase
transition (see \cite{ckn93} for reviews). The order parameter used to
measure the strength of the elecroweak phase transition is the mean value of
the Higgs field in the broken symmetry phase at the critical temperature, $
v\left( T_{c}\right) =\langle \phi \rangle _{T_{c}}$.

In the standard mechanism for electroweak baryogenesis, which assumes a
first-order phase transition, the non-equilibrium condition acts in two
different ways. The first of them is through the expansion of bubbles of the
stable phase, which combined with $CP$ violation inside the walls of bubbles
produce non-equilibrium particle densities, which then give rise to the
baryon asymmetry. The second effect of the departure from equilibrium is
connected to the baryon number violation condition: baryon number violating
processes must be turned off before the system reaches thermal equilibrium
in order to avoid the washout of the generated BAU.

In the expansion of bubbles an asymmetry between left handed quarks and
their antiparticles (and an opposite right handed asymmetry) is built up in
front of the moving walls. This asymmetry biases the anomalous, baryon
number violating sphaleron processes present in the high temperature phase.
Thus, bubble walls must have a nonvanishing velocity to generate a net
baryon asymmetry. On the other hand, the left-handed asymmetry injected in
front of the wall needs some time to diffuse and bias the sphaleron
processes before the reflected particles are re-caught by the wall. If the
wall moves too fast, there won't be enough time for sphalerons to produce
baryons. As a consequence, the generated baryon asymmetry has a peak at some
small velocity which depends on the time scales associated to particle
diffusion and baryon number violation \cite{nkc92,h95,ck00}. In particular,
it was shown in Ref.~\cite{ck00} that for the MSSM the peak is at $
v_{w}=0.01-0.03$.

The chiral quark asymmetry generated in front of the bubble wall depends
also on the wall thickness. If the wall width is small compared to the
inverse temperature, the $CP$-violating reflection of particles by the
bubble wall can be treated quantum mechanically \cite{ckn91}. Otherwise, the
effect of $CP$ violation acts as a classical force on the quarks as they
pass through the wall \cite{jpt95}. This is the case in the electroweak
phase transition, since the wall width has been estimated to be $l_{w}\sim
10T^{-1}$ \cite{cm98,dlhll92,mqs98}. The final baryon asymmetry will thus
also depend on $l_{w}$, resulting a larger asymmetry for thinner walls.

Regarding the avoiding of washout, sphaleron processes must be suppressed in
the broken phase so that baryons that are created in front of the moving
wall are not erased after entering the bubble. This requirement imposes a
constraint on the sphaleron energy which can be expressed as the well known
condition for the Higgs mean value \cite{s86}
\begin{equation}
\frac{v\left( T_{c}\right) }{T_{c}}\gtrsim 1\ .  \label{washout}
\end{equation}
This condition gives a severe constraint for the theory of the electroweak
interactions. Indeed, the minimal SM is unable to explain the observed BAU,
since in this model the phase transition is not of the first order, but only
a smooth crossover \cite{bf94}. Furthermore, electroweak baryogenesis
constrains the Minimal Supersymmetric Standard Model (MSSM) to a small
region of parameter space, where the Higgs mass is less than about $105GeV$
and the stop mass is less than the top one \cite{cqw98,cm98}.

From the above it becomes apparent that phase transition dynamics plays a
relevant role in electroweak baryogenesis. Being the baryon production very
sensitive to the bubble wall width and velocity, and to the Higgs mean
value, it is important to determine as accurately as possible the values of
these parameters at the actual temperature of the phase transition. Indeed,
all of these parameters vary significantly in the temperature interval in
which the first-order phase transition can take place, that is to say,
between the critical temperature $T_{c}$ at which the two phases have the
same free energy and the temperature $T_{0}$ at which the barrier between
the two minima of free energy disappears. The exact temperature $T_{t}$ of
the transition depends on the evolution of the bubbles after they are
nucleated, which in turn depends on the viscosity of the plasma.

In this paper we will analyze in detail the dependence on temperature of the
parameters $l_{w}$, $v_{w}$, and $v\left( T\right) /T$, and we will
determine the relative position of the temperature $T_{t}$ in the interval $
T_{0}-T_{c}$. We will do this computation for different values of the
parameters of our model, placing emphasis on those that adequate to the
context of the MSSM. We will also discuss several aspects of the phase
transition dynamics, such as transitory states in the evolution of the Higgs
field, or the possibility that the transition occurs near the temperature $
T_{0}$, where the mechanism of baryogenesis would be different from the
usual one.

The plan is the following. In the next section we introduce our model for
the first-order elecroweak phase transition, and calculate the shape of the
nucleated bubbles as a function of temperature. In section \ref{growth} we
study the subsequent evolution of the bubbles in the hot plasma. In section
\ref{temp} we determine a temperature interval outside which the phase
transition cannot occur, independently of the bubble evolution. Finally, in
section \ref{develop} we compute the temperature and duration of the
transition and discuss the implications for baryogenesis. In the appendix we
show some details of the numerical calculation.

\section{The phase transition \label{pht}}

The electroweak phase transition takes place when the expectation value of
the Higgs field passes from its high temperature value $\langle \phi \rangle
=0$ to its nonzero value in the low temperature broken phase. We will use a
simple model for the phase transition, which is nevertheless representative
of the electroweak theory at high temperature.

\subsection{The effective potential}

We will assume that the free energy has the well known general form \cite
{dlhll92,ah92,eqz93}
\begin{equation}
V\left( \phi ,T\right) =D\left( T^{2}-T_{0}^{2}\right) \phi ^{2}-ET\phi
^{3}+\lambda \phi ^{4}\ ,  \label{pot}
\end{equation}
which will be suitable for our analysis, since it contains the essential
features of the first order phase transition. Moreover, it can be a very
good approximation to the actual effective potential. This is for instance
the form of the perturbative high temperature effective potential in the SM
\cite{dlhll92,ah92}. Even in the MSSM, which has two Higgs doublets, the
free energy takes the SM-like form (\ref{pot}) in the limit in which the
pseudoscalar particle of the Higgs sector is heavy ($m_{A}\gg T_{c}$) \cite
{eqz93}.

All the parameters in Eq.~(\ref{pot}) depend on the particle content of the
theory. Parameter $D$ contains contributions from all the particles that
acquire their masses through the Higgs mechanism. These contributions are of
the form $m_{i}^{2}/v^{2}$, where $m_{i}$ is the zero-temperature mass of
particle $i$ and $v$ is the zero-temperature vacuum expectation value of the
Higgs field. Parameter $E$ has only boson contributions, of the form $
m_{i}^{3}/v^{3}$. In the SM we have $D\sim 10^{-1}$, $E\sim 10^{-2}$, while
in the MSSM, due to the larger particle zoo, $D$ and $E$ can be more than an
order of magnitude larger than in the SM. Parameter $\lambda $ is in general
temperature dependent, but it is almost constant in the range of
temperatures in which the phase transition can take place. This parameter is
very sensitive to the Higgs mass and for our discussion we will assume it to
be given parametrically by $m_{H}\sim \lambda v^{2}$.

The cubic term in $V\left( \phi ,T\right) $ is responsible for the
first-order feature of the phase transition, by causing the coexistence of
two minima separated by a barrier. Hence, the strength of the transition
depends on the value of parameter $E$. At high temperature the global
minimum of the potential is at $\phi =0$. At the critical temperature
\begin{equation}
T_{c}=\frac{T_{0}}{\sqrt{1-\frac{E^{2}}{\lambda D}}}  \label{tc}
\end{equation}
the two minima become degenerate, and below this temperature the stable
minimum of $V$ is at
\begin{equation}
v\left( T\right) =\frac{3ET}{2\lambda }\left( 1+\sqrt{1-\frac{8}{9}\frac{
\lambda D}{E^{2}}\left( 1-\frac{T_{0}^{2}}{T^{2}}\right) }\right) \ .
\label{fimin}
\end{equation}
At temperature $T_{0}$ the barrier between minima disappears, and $\phi =0$
becomes a maximum of the potential. The exact value of $T_{0}$ depends on
the parameters of the theory, but we can assume it to be roughly $\sim
100GeV $ since the dynamics of the phase transition is not very sensitive to
the value of this constant. The number $E^{2}/\lambda D$ is in general
small, and the difference between $T_{c}$ and $T_{0}$ is $\Delta T\lesssim
10^{-2}T_{c}$. However, as we shall see, things change rapidly as the
temperature falls from $T_{c}$ to $T_{0}$. Hence it proves useful to define
a dimensionless variable
\begin{equation}
\varepsilon \equiv \frac{T_{c}^{2}-T^{2}}{T_{c}^{2}-T_{0}^{2}}\simeq \frac{
T_{c}-T}{T_{c}-T_{0}}\ ,  \label{epsilon}
\end{equation}
which goes from $0$ to $1$ as $T$ runs between $T_{c}$ and $T_{0}$.

At the critical temperature, Eq.~(\ref{fimin}) gives $v\left( T_{c}\right)
=2ET_{c}/\lambda $. Inserting this value into the condition for avoiding the
washout of the baryon asymmetry [Eq.~(\ref{washout})], gives the constraint $
\lambda \lesssim 2E$. In the SM this results in the bound on the Higgs mass $
m_{H}\lesssim 40GeV$, well below the experimental bound $m_{H}^{\exp
}\gtrsim 95GeV$. In the MSSM, the one loop stop contribution increases the
value of parameter $E$, but this contribution is limited by the danger of
inducing charge and color breaking minima. As a consequence, the bound on $
m_{H}$ can be shifted only to $\sim 80GeV$, provided that the light stop
mass is in the range $100GeV\lesssim m_{\tilde{t}}\lesssim m_{t}$ \cite
{cqrvw97}. It turns out that two loop corrections produce an enhancement of
the strength of the phase transition, leading to an upper bound on the Higgs
mass of $\sim 105GeV$ \cite{cqw98,cm98}.

Although Eq.~(\ref{washout}) is written in terms of the critical temperature
$T_{c}$, this condition is somewhat inexact, since at $T_{c}$ the nucleation
of bubbles has not yet begun. On the other hand, the phase transition must
be completed when the temperature of the Universe is still above $T_{0}$,
because bubbles of the true phase are copiously produced as the barrier
between minima gets small and disappears. Hence, the transition takes place
at an intermediate temperature $T_{0}<T_{t}<T_{c}$ with
\begin{equation}
\frac{v\left( T_{c}\right) }{T_{c}}<\frac{v\left( T_{t}\right) }{T_{t}}<
\frac{v\left( T_{0}\right) }{T_{0}}\ .
\end{equation}
The temperature $T_{t}$ is usually estimated by the condition $S_{3}\left(
T_{t}\right) /T_{t}\sim 130-140$, where $S_{3}$ is the fluctuation in free
energy which is necessary for bubble formation \cite
{dlhll92,ah92,eqz93,cqrvw97,mstv91}. However, as was stressed in Ref.~\cite
{ah92}, a more careful determination of the temperature $T_{t}$ is
important, because the rate of anomalous baryon number violation is an
exponentially sensitive function of $v\left( T\right) /T$. In the effective
potential (\ref{pot}), $v\left( T\right) /T$ varies from $2E/\lambda $ at
temperature $T_{c}$ to $3E/\lambda $ at $T_{0}$. This variation by a factor
of $3/2$ gives an uncertainty of a 50\% in the bound on the Higgs mass
coming from Eq.~(\ref{washout}). Elimination of this uncertainty is thus
essential given the current experimental bounds on $m_{H}$. So the correct
condition for avoiding the washout of the baryon asymmetry would be,
according to Eq.~(\ref{fimin}),
\begin{equation}
\lambda <p\left( T\right) E\ ,  \label{lambdae}
\end{equation}
where $p\left( T\right) $ is a number between $2$ and $3$, to be evaluated
at the temperature of the phase transition, $T_{t}$.

\subsection{Bubble nucleation}

At a temperature below $T_{c}$, bubbles of the low temperature phase begin
to nucleate. The thermal tunneling probability (per unit volume and time)
for bubble nucleation is \cite{l77}
\begin{equation}
\Gamma \sim A\left( T\right) e^{-S_{3}/T}\ .  \label{gamma}
\end{equation}
The prefactor $A\left( T\right) $ is roughly of order $T^{4}$ (so we will
set $A\left( T\right) =T^{4}$), and $S_{3}\left( T\right) $ is the
three-dimensional instanton action, which coincides with the free energy of
the nucleated bubble. At high temperature the bounce solution is $O\left(
3\right) $ symmetric, and the corresponding action takes the simple form
\begin{equation}
S_{3}=4\pi \int_{0}^{\infty }r^{2}dr\left[ \frac{1}{2}\left( \frac{d\phi }{
dr }\right) ^{2}+V\left( \phi \left( r\right) ,T\right) \right] \ ,
\end{equation}
where $r^{2}=\mathbf{x}^{2}$, and the equation for the configuration of the
bubble is
\begin{equation}
\frac{d^{2}\phi }{dr^{2}}+\frac{2}{r}\frac{d\phi }{dr}=V^{\prime }\left(
\phi \right) \ ,  \label{ecfi}
\end{equation}
with the boundary conditions $\phi \left( r=\infty \right) =0$ and $\frac{
d\phi }{dr}\left( r=0\right) =0$.

For temperatures close to $T_{c}$, the width of the bubble wall at the
moment of formation is much smaller than its radius, and $S_{3}$ can be
expressed as a function of the bubble radius $r_{0}$, the energy difference $
\Delta V$ between the two minima of the potential, and the bubble wall
surface energy $\sigma $ \cite{l77}. The radius of the critical bubble to be
nucleated can thus be obtained by finding an extremum of $S_{3}\left(
r_{0}\right) $. A similar approximation can be used to estimate the radius
of a thick-walled bubble (i.e., outside the range of validity of the thin
wall approximation) \cite{ah92}. However, as was pointed out in Ref.~\cite
{dlhll92}, one must be very careful when evaluating $S_{3}$ in this way.
Since the solution of Eq.~(\ref{ecfi}) is a \emph{maximum} of the action,
the corresponding value of $S_{3}$ will be higher than the action of any
approximate solution. As a consequence one can strongly overestimate the
tunneling probability, leading to a sooner completion of the phase
transition.

For the potential (\ref{pot}), Eq.~(\ref{ecfi}) can be solved numerically
and then integrated to obtain the action $S_{3}$, which can be expressed as
a function of a dimensionless parameter $\alpha =\frac{\lambda D}{E^{2}}
\left( \frac{T^{2}-T_{0}^{2}}{T^{2}}\right) $,
\begin{equation}
\frac{S_{3}}{T}=13.72\frac{E}{\lambda ^{3/2}}\ \alpha ^{3/2}f\left( \alpha
\right) \ .  \label{s3}
\end{equation}
Dine \textit{et al.~}\cite{dlhll92} have found a useful analytical
approximation to their numerical result for the function $f\left( \alpha
\right) $,
\begin{equation}
f\left( \alpha \right) =1+\frac{\alpha }{4}\left( 1+\frac{2.4}{1-\alpha }+
\frac{0.26}{\left( 1-\alpha \right) ^{2}}\right) \ ,  \label{f}
\end{equation}
with an accuracy of about 2\% in the interval\footnote{It
is easy to see from Eq.~(\ref{tc}) that $\alpha \left( Tc\right) =1$, $
\alpha \left( T_{0}\right) =0$, and $\alpha \simeq 1-\varepsilon $ in the
whole interval.} $0<\alpha <1$. Using the same procedure we have calculated
the radius $r_{0}$ of the nucleated bubble, the thickness $l_{w}$ of its
wall, and the value of the Higgs field inside it, $\phi _{0}$, as functions
of $\alpha $. The calculation is described in the Appendix. Notice that the
dimensionless function $p\left( T\right) $ defined by Eqs. (\ref{fimin}) and
(\ref{lambdae}) has also a simple expression as a function of $\alpha $,
\begin{equation}
p\left( \alpha \right) =\frac{3}{2}\left( 1+\sqrt{1-\frac{8}{9}\alpha }
\right) \ .  \label{palfa}
\end{equation}

\begin{figure}[tbh]
\centering
\epsfysize=5cm \leavevmode
\epsfbox{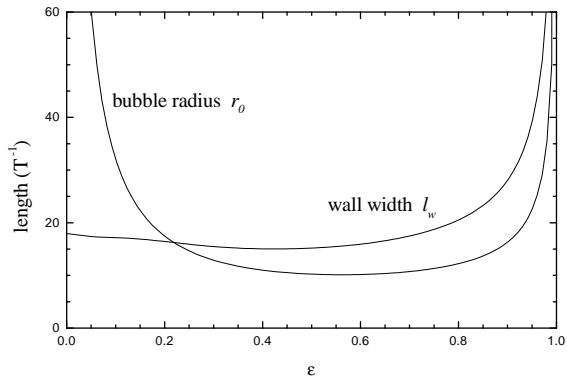}
\caption{Bubble radius and wall thickness of the nucleated bubbles vs the
relative messure of the temperature, $\varepsilon $, for $D=1$, $E=0.06$,
and $\lambda =2E$.}
\label{figrl}
\end{figure}

\begin{figure}[tbh]
\centering
\epsfysize=5cm \leavevmode
\epsfbox{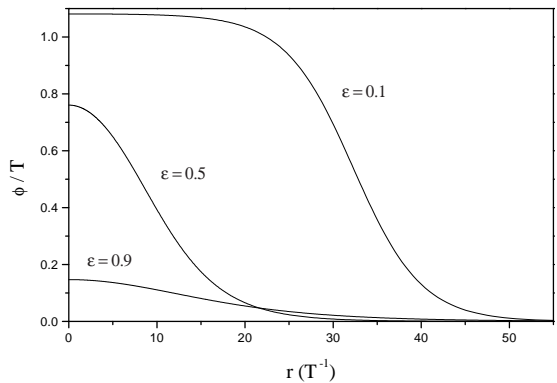}
\caption{Shape of the critical bubble at different temperatures.}
\label{shape}
\end{figure}

\begin{figure}[tbh]
\centering
\epsfysize=5cm \leavevmode
\epsfbox{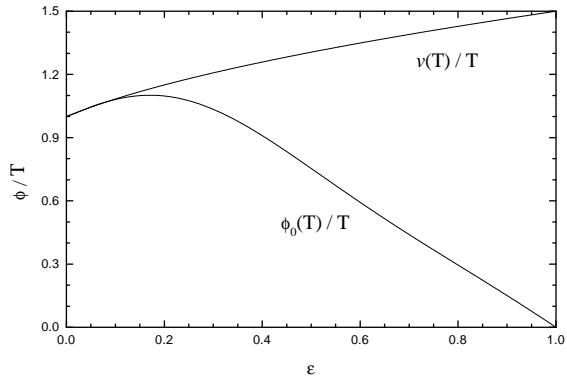}
\caption{The minimum of the potential $v\left( T\right) $ and the value of $
\phi $ inside the nucleated bubble, $\phi _{0}\left( T\right) \equiv \phi
\left( r=0\right) $.}
\label{figfi}
\end{figure}

Figure~\ref{figrl} illustrates the behavior of $r_{0}\left( \varepsilon
\right) $ and $l_{w}\left( \varepsilon \right) $. Near the critical
temperature ($\varepsilon =0$) the bubbles that nucleate are thin-walled ($
l_{w}\ll r_{0}$). This is because the energy difference between the two
minima of the potential is much smaller than the energy barrier between
them. In such a case the radius must be very large, so that the negative
potential energy inside the volume of the bubble is able to compete with the
positive surface energy. As $\varepsilon $ increases, the wall width becomes
of the same order of the radius and the thin wall approximation breaks down.
For $\varepsilon \rightarrow 1$ the size of the critical bubbles diverges.
This is because the width of the bubble wall is associated to the
correlation length, $l_{w}\sim \xi \left( T\right) \sim m\left( T\right)
^{-1}$, with $m\left( T\right) =2D\left( T^{2}-T_{0}^{2}\right) $. This
divergence is easily understood by looking at the shape of the bubbles.
Figure~\ref{shape} shows the bubble profile at a temperature near $T_{c}$ ($
\varepsilon =0.1$), a temperature near $T_{0}$ ($\varepsilon =0.9$), and at
an intermediate temperature ($\varepsilon =0.5$). Note that the value of the
Higgs field inside the bubble decreases with $\varepsilon $.
In Fig.~\ref{figfi} we have plotted this value together with the minimum of
the potential $v\left( T\right) $ for comparison. At $\varepsilon =0$ the
value of $\phi _{0}$ coincides with that of $v\left( T_{c}\right) $.
However, as the temperature decreases from the critical one, this value
moves away from the minimum of $V\left( \phi \right) $ and goes to zero.
This behavior is due the decreasing of the height of the barrier and the
increasing in the energy difference between minima. In this case the value
of $\phi \left( x\right) $ inside the bubble needs not anymore to be exactly
at the minimum of the potential to get enough volume energy. Moreover, as
the barrier between minima disappears, it becomes easier to form a large
bubble with a small value of $\phi $ inside it. Therefore, near the
temperature $T_{0}$, bubbles with $r_{0}\rightarrow \infty $ and $\phi
_{0}\rightarrow 0$ will be formed.

Consequently, if for any reason the transition does not occur until the
temperature is close to $T_{0}$, then it will tend to proceed by a
homogeneous growing of the order parameter $\phi $, instead of by the
nucleation and expansion of bubbles. This is not surprising since, without a
barrier, one expects the field to ``roll down'' homogeneously towards the
minimum of the effective potential \cite{ftmm99}. There would still be some
baryon number generation in such a scenario, due to the coherent variation
of the $CP$ violating field $\phi \left( t\right) $, which would produce an
effective chemical potential for baryon number. However, this would be a
kind of local baryogenesis \cite{ckn93} mechanism, and hence it would lack
the enhancement due to diffusion, which characterizes the usual non-local
electroweak baryogenesis. Furthermore, after rolling down the field will
start to oscillate around the minimum of the potential until all its energy
is transferred to the plasma. This oscillations can further spoil the
generated baryon asymmetry, depending on their amplitude. This effect would
be avoided in a phase transition at a higher scale, where the motion of $
\phi $ would be strongly damped by the expansion of the Universe apart from
the damping due to the plasma. Nevertheless, we shall see in section \ref
{temp} that the phase transition is unlikely to occur beyond $\varepsilon
=0.5$, so the value of $\phi $ inside the bubbles at the moment of
nucleation will not be far from $v\left( T\right) $.

We also see from Fig.~\ref{figrl} that the wall thickness is almost constant
up to that value of $\varepsilon $, so we can anticipate that the width $
l_{w}$, (which enters the generated baryon number), will depend weakly on
the temperature of the phase transition.

\section{The evolution of bubbles \label{growth}}

After a bubble is formed, it starts to grow due to the pressure difference
between the phases separated by the bubble wall, $\Delta V=V\left(
v,T\right) -V\left( 0,T\right) $. After a short acceleration stage, the wall
reaches a terminal velocity which depends on its interaction with the
plasma. In this section we will analyze the evolution of bubbles from the
moment they are nucleated until they fill the Universe

\subsection{Bubble wall propagation}

If the time it takes a bubble wall to accelerate up to terminal velocity is
much smaller than the time it propagates before colliding with other
bubbles, then the initial acceleration stage can be ignored. In order to
check that this is the case, we can estimate the time needed for the wall to
reach terminal velocity from the equation of motion for $\phi $ in the
plasma, Eq.~(\ref{ew3}) below. If the field is strongly damped the wall will
reach terminal velocity faster than without damping, so using
Eq.~(\ref{ew3})
without the damping term one can estimate the maximum acceleration time
\cite{h95}. In such a case the wall reaches the speed of light in a time $
\tau \sim v\left( T\right) ^{2}/\left( l_{w}V\left( v\left( T\right)
,T\right) \right) $. For the values of the parameters of $V\left( \phi
,T\right) $ that we will consider, this gives $\tau \sim \left(
10^{2}-10^{4}\right) T^{-1}$.

In order to compare this time with the typical times associated with the
phase transition, we need the relation between the temperature of the
Universe and time. At the radiation-dominated era it is given by
\begin{equation}
t=\frac{\xi M_{Pl}}{T^{2}}\ ,  \label{ttrad}
\end{equation}
where $M_{Pl}=1.22\times 10^{19}GeV$ is the Plank mass, and $\xi $ is a
factor that depends on the effective degrees of freedom of the plasma. Near
the temperature of the electroweak phase transition, $\xi \simeq 1/34$. The
time elapsed between $T_{c}$ and $T_{0}$, for instance, is
\begin{equation}
\Delta t_{T_{c}-T_{0}}\sim \frac{\xi M_{Pl}}{T_{c}}\frac{T_{c}-T_{0}}{T_{c}}
T_{c}^{-1}\sim 10^{13}T^{-1}\ ,  \label{dtc0}
\end{equation}
which is many orders of magnitude larger than $\tau $. However, to check if
the short acceleration assumption is correct we must compare $\tau $ with
the time in which a typical bubble propagates until all space is filled by
broken symmetry bubbles. This time is of the order of the duration of the
phase transition $\Delta t_{t}$, which is less than $\Delta
t_{T_{c}-T_{0}}$. We will see in section \ref{develop} that $\Delta
T_{t}\simeq 10^{-3}\left( T_{c}-T_{0}\right) $, so $\Delta t_{t}\sim
10^{10}T^{-1}$. The assumption is thus valid.

There is another transitory stage, associated with the growing of $\phi $ in
the interior of the bubble. As seen in Fig.~\ref{figfi}, bubbles nucleate
with a value of the field inside them, $\phi _{0}\left( T\right) $, that is
less than that of the minimum of the potential $v\left( T\right) $. This
fact can be important for the bubble wall propagation, since the pressure
difference $V\left( \phi _{0},T\right) -V\left( 0,T\right) $ that pushes the
wall would then be less than $V\left( v,T\right) -V\left( 0,T\right) $.
Again, we can use Eq.~(\ref{ew3}), this time discarding all spatial
derivatives, to calculate the time in which $\phi \left( r=0\right) $ grows
from $\phi _{0}$ to $v$. Neglecting the damping term we obtain $\tau \sim
\sqrt{\Delta \phi /V^{\prime }\left( \phi \right) }$. For $\Delta \phi
\simeq \phi _{0}\simeq v/2$ this gives $\tau \sim 10T^{-1}$. With a damping
term it will take the field a longer time to get to $v$. Assuming a typical
damping $\eta \sim 100$ (see below), we obtain $\tau \sim \Delta \phi \eta
/V^{\prime }\left( \phi \right) \sim \left( 10^{4}-10^{5}\right) T^{-1}$,
which is again much less than $\Delta t_{t}$.

We can therefore safely assume that $\phi \left( r=0\right) =v\left(
T\right) $, and that bubble walls move with constant velocity through the
plasma. The friction on the bubble wall depends on the departure from
equilibrium of the particle distributions near the wall \cite
{dlhll92,t92,k92}. Particles which interact with the Higgs field in such a
way that their masses change across the bubble wall profile will feel a
force as they are caught by the propagating wall. Integration of these
forces weighted with the particle distributions gives the total force
exerted on the wall by the plasma. When equilibrium distributions
appropriate for the local value of $\phi $ are considered, this force only
accounts for the finite temperature part of $\Delta V$, and does not depend
on the wall velocity. Nevertheless, a departure from equilibrium is in fact
produced by the motion of the wall. Particles that, for energetic reasons,
cannot penetrate the wall, are reflected in front of it. These particles
slow down the wall by giving momentum to it, and this effect becomes more
important the faster the wall moves. Expanding the non-equilibrium particle
distributions to first order in the wall velocity, the force gets a
contribution proportional to the wall velocity \cite{dlhll92}. The terminal
velocity is thus obtained by equating this frictional force with $\Delta V$.

The propagation of the bubble wall can also be affected by hydrodynamics
(see for instance \cite{h95,ikkl94,eikr92}). Although hydrodynamic
considerations are important for large velocities, if the wall velocity is
small the only important effect is a global reheating due to the latent heat
released by the expanding bubbles \cite{h95}. Indeed, when the velocity of
the bubble wall is less than the speed of sound in the relativistic plasma $
c_{s}=\sqrt{1/3}$, it propagates as a deflagration front. This means that a
shock front precedes the wall, with a velocity $v_{sh}>c_{s}$. Hence, the
latent heat is transmitted away from the wall. If the wall velocity is small
enough, the latent heat has time to distribute throughout space, causing a
homogeneous reheating of the plasma as the wall propagates.

There exist other effects that can influence the growth of bubbles. An
example is that of strong magnetic fields that can be formed before \cite
{tw88} or during \cite{e98} the electroweak phase transition. The presence
of these fields changes the free energy difference between phases, thus
affecting the temperature and strength of the first order phase transition
\cite{gs98}. In the case that the magnetic fields are generated during the
electroweak phase transition, they can delay the completion of the latter by
strongly affecting the evolution of bubbles \cite{ftmm99}.

\subsection{The wall velocity}

The equation of motion for the bubble wall in the hot plasma can be derived
by energy conservation considerations \cite{dlhll92,t92,k92} ,
\begin{equation}
\partial _{\mu }\partial ^{\mu }\phi +\frac{\partial V\left( \phi ,T\right)
}{\partial \phi }+\sum_{i}\frac{\partial m_{i}^{2}}{\partial \phi }\int
\frac{d^{3}p}{\left( 2\pi \right) ^{3}2E_{i}}\delta f_{i}=0\ ,  \label{ew1}
\end{equation}
where $\delta f_{i}$ is the deviation from the equilibrium distribution
function for particle species $i$ in the heat bath. The last term can be
expressed \cite{h95,ikkl94} as a frictional damping term proportional to $
d\phi /dt$. Due to Lorentz invariance this term must be of the form $u^{\mu
}\partial _{\mu }\phi $, where $u_{\mu }$ is the four-velocity of the
plasma. Eq.~(\ref{ew1}) then becomes
\begin{equation}
\partial _{\mu }\partial ^{\mu }\phi +V^{\prime }\left( \phi \right) +\left(
\eta T\right) u^{\mu }\partial _{\mu }\phi =0\ ,  \label{ew3}
\end{equation}
where $\eta $ is a dimensionless damping coefficient that depends on the
viscosity of the medium, and is obtained by calculating the deviations $
\delta f_{i}$ near the wall.

The calculation of the friction is usually treated with kinetic theory \cite
{dlhll92,t92,k92}, giving for the Standard Model roughly $\eta \sim 1$.
However, it was noticed \cite{m00} that infrared boson excitations which are
treated classically \cite{mt97}, increase the value of $\eta $ by an order
of magnitude. In the MSSM there is an additional enhancement of the friction
due to the larger particle content of the model. In Ref. \cite{js00} the
contribution of a light right-handed stop has been calculated, which added
to the contributions of tops and $W$ bosons give a lower bound on the
friction coefficient of $\eta \simeq 70$.

We will not take into account other effects such as the influence of
magnetic fields on the bubble growth. Instead, we will consider values of
the friction beyond $\eta \sim 100$ to allow for the possibility of
additional damping, and we will assume that that kind of effects can be
parametrized in this way. Accordingly, we will let $\eta $ vary in the range
$1<\eta <1000$.

The nucleated bubbles grow rapidly to a size large enough so that we can go
to 1+1 dimensions. Boosting to a frame that moves with the wall and assuming
stationary motion in the $x$ direction we then have
\begin{equation}
\frac{d^{2}\phi }{dx^{2}}=V^{\prime }\left( \phi \right) -\eta Tv_{w}\gamma
\frac{d\phi }{dx}\ ,  \label{ew2}
\end{equation}
where $\gamma =1/\sqrt{1-v^{2}}$. Multiplying both sides by $d\phi /dx$ and
integrating over $-\infty <x<\infty $, we obtain a formula for the wall
velocity,
\begin{equation}
v_{w}\gamma =\frac{1}{\eta T}\frac{\Delta V}{\sigma }\ ,
\end{equation}
where $\Delta V\left( T\right) $ is the free energy difference between the
two phases and $\sigma \left( T\right) $ is the surface tension of the wall,
\begin{equation}
\sigma \left( T\right) =\int_{-\infty }^{+\infty }\left( \frac{d\phi }{dx}
\right) ^{2}dx\ .  \label{sigma}
\end{equation}
We have assumed here that temperature is constant across the wall. This is
right if the wall velocity is small enough that the latent heat it releases
has time to be uniformly distributed throughout space \cite{h95}. If we use
the well known kink approximation for the wall profile (see for instance
\cite{j99})
\begin{equation}
\phi \left( x\right) =\frac{v\left( T\right) }{2}\left[ 1+\tanh \left( \frac{
x}{l_{w}}\right) \right] \ ,
\end{equation}
then the surface energy (\ref{sigma}) is $\sigma =v^{2}/3l_{w}$, and the
wall velocity is given by
\begin{equation}
v_{w}\gamma =\frac{3l_{w}\Delta V}{T\eta v^{2}}\ .  \label{veloc}
\end{equation}

We see from Eq.~(\ref{veloc}) that $v_{w}$ increases with $l_{w}$. This is
because the thicker the bubble wall, the slower will be the variation of $
\phi $ as the wall sweeps through a given point in the plasma, and
consequently, the smaller the disturbance of the plasma near the wall.
Hence, thicker walls experience a smaller friction force from the medium.
Since $l_{w}$ diverges at $T_{0}$, one can infer that the dependence of the
velocity with the wall width gives an important variation of $v_{w}$ with
temperature. In the literature, however, $l_{w}$ is usually evaluated near
the critical temperature. For instance, it is often roughly approximated by
the correlation length $m\left( T_{c}\right) ^{-1}$ \cite{h95,m00}, or it is
taken from the wall profile at the temperature at which $S_{3}\left(
T\right) /T\sim 130-140$ \cite{cm98,dlhll92,mqs98,js00}. This gives $
l_{w}\sim \left( 10-40\right) T^{-1}$ for the electroweak bubble. We can see
from Fig.~\ref{figrl} that these approximations are good as long as the
phase transition completes before the temperature gets close to $T_{0}$,
since $l_{w}$ is almost constant until $\varepsilon \simeq 0.8$. As we will
see in the next section, the phase transition hardly occurs beyond that
value of $\varepsilon $.

\begin{figure}[tbh]
\centering
\epsfysize=5cm \leavevmode
\epsfbox{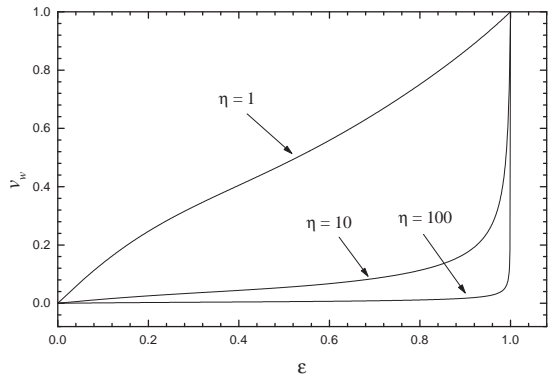}
\caption{Bubble wall velocity as a function of $\varepsilon $, for three
different values of the damping coefficient $\eta $.}
\label{vw}
\end{figure}

Figure~\ref{vw} shows the terminal velocity of the bubble wall as a function
of $\varepsilon $. In the interval $0<\varepsilon <1$ it varies from $v_{w}=0
$ (since $\Delta V=0$ at the critical temperature) to $v_{w}=1$ (since $
l_{w}\rightarrow \infty $ for $\varepsilon \rightarrow 1$). Note, however,
that at the beginning of the interval the wall velocity grows rapidly from $0
$ to some value proportional to $\eta ^{-1}$, then it becomes rather
insensitive to $\varepsilon $, and finally it grows again up to $v_{w}=1$.
As we shall see, the electroweak phase transition generally occurs at some
intermediate $\varepsilon $ so, for fixed $\eta $, the wall velocity will
not be very sensitive to the temperature of the transition.

\subsection{Fraction of space occupied by bubbles}

Once a bubble has nucleated, its wall rapidly reaches the terminal velocity $
v_{w}$, which depends on temperature. Due to the large time-temperature
relationship (\ref{ttrad}), $v_{w}$ varies slowly with time, so the
evolution of the bubble radius can be approximated by
\begin{equation}
r\left( t^{\prime },t\right) =r_{0}\left( t^{\prime }\right) +v_{w}\left(
t^{\prime }\right) \left( t-t^{\prime }\right)   \label{radio}
\end{equation}
where $r_{0}\left( t^{\prime }\right) $ is the radius of the bubble at the
moment $t^{\prime }$ of its formation\footnote{By
the epoch of the electroweak phase transition, the variation of length
scales due to the expansion of the Universe can be neglected during the
short period between $T_{c}$ and $T_{0}$.}. The number of bubbles created
per unit volume between $t^{\prime }$ and $t^{\prime }+dt^{\prime }$ is $
dn_{b}\left( t^{\prime }\right) =\Gamma \left( t^{\prime }\right) dt^{\prime
}$. Then at time $t$ the total volume of bubbles that have nucleated at
previous times $t_{c}<t^{\prime }<t$ is (per unit volume of space) $
\int_{t_{c}}^{t}\Gamma \left( t^{\prime }\right) \left( 4\pi /3\right)
r\left( t^{\prime },t\right) ^{3}dt^{\prime }$. This formula, however,
contains multiple counting of volume where bubbles overlap. A more exact
expression was obtained in Ref.~\cite{gw81} for the fraction of space $
f\left( t\right) $ that is still in the false vacuum,
\begin{equation}
f\left( t\right) =\exp \left\{ -\frac{4\pi }{3}\int_{t_{c}}^{t}\Gamma \left(
t^{\prime }\right) \left[ r_{0}\left( t^{\prime }\right) +v_{w}\left(
t^{\prime }\right) \left( t-t^{\prime }\right) \right] ^{3}dt^{\prime
}\right\} \ .  \label{frac}
\end{equation}
The phase transition then completes when $f\left( t\right) $ falls to zero.
All the quantities appearing in the integral are in fact functions of
temperature, so the use of the time-temperature relation (\ref{ttrad}) will
be needed in the computation.

Some reheating will occur at the end of the transition due to collisions of
bubble walls. However, for small wall velocities this effect should not be
important. As we have already mentioned, the main consequence of the release
of energy by slowly-expanding bubbles is a global reheating \emph{during}
the bubble expansion stage, which would manifest itself by changing the
time-temperature relation.
Taking into account this effect would require solving the two coupled
equations for $T\left( t \right)$ and $f\left( t \right)$, which is outside
the scope of the present paper. Before
going on, however, we would like to discuss the relevance of such an
effect.
As pointed out in Ref.~\cite{h95}, one can get an idea of how
important the reheating is by comparing the latent heat
\begin{equation}
L\equiv \left. T\frac{\partial V\left( \phi ,T\right) }{\partial T}\right|
_{v,T_{c}}=8\frac{DE^{2}}{\lambda ^{2}}T_{0}^{4}
\end{equation}
with the energy needed to bring the plasma back up to $T_{c}$ from $T_{t}$,
\begin{equation}
\rho \left( T_{c}\right) -\rho \left( T_{t}\right) =a\left(
T_{c}^{4}-T_{t}^{4}\right) \ ,
\end{equation}
where the factor $a$ depends on the number of effectively massless degrees
of freedom. At the time of the electroweak phase transition $a\simeq 35$.
Since $T_{c}-T_{t}\lesssim T_{c}-T_{0}$, the energy density difference will
be $\Delta \rho \lesssim T_{c}^{4}$. For SM values of $D$ and $E$, the
latent heat is smaller than $\Delta \rho $ by one or two orders of
magnitude, depending on the value of $\lambda $. For MSSM values of the
parameters, instead, $L$ can be of the same order of magnitude of $\Delta
\rho $. In such a case the latent heat released by the bubbles could reheat
the plasma back up to temperatures near $T_{c}$. So we must remark that this
effect might be important for baryogenesis. Due to the small velocity of the
bubble walls, the gradually released latent heat will prevent the
temperature of the bath to decrease, rather than reheating the plasma at the
end of the transition. If the temperature stays close to $T_{c}$, the
pressure difference $\Delta V$ across the bubble walls will be small, and
hence the wall velocity can decrease considerably from the current
estimations \cite{h95}.

\section{The temperature of the phase transition \label{temp}}

The progress of the phase transition is determined by the fraction of space $
f\left( T\right) $ that is still in the symmetric phase, given by Eq.~(\ref
{frac}). In the next section we will calculate the temperature $T_{t}$ of
the transition and its duration $\Delta T_{t}$ for different values of the
friction. However, since the exact evolution of the bubble wall velocity is
not known, it will be convenient to begin our analysis by putting reliable
constraints on $T_{t}$ inside the interval $T_{0}-T_{c}$.

We will consider two sets of values for the parameters $E$ and $D$ of the
potential: those corresponding to the SM, $E\simeq 6\times 10^{-3}$, $
D\simeq 0.2$; and MSSM-like values, which for definiteness we set $E\simeq
6\times 10^{-2}$, $D\simeq 1$. In both cases we will take $\lambda /E=2$ or $
3$, which correspond respectively to $v\left( T_{c}\right) /T_{c}=1$ or $
v\left( T_{0}\right) /T_{0}=1$. Although the SM with these values of the
parameters is not a realistic model, for the sake of comparison it is useful
to consider such a kind of parameters besides those corresponding to the
MSSM, in order to simulate a different phase transition with the same simple
form of the effective potential. Regarding the MSSM-like values, it is
worthy to remind that our model Eq.~(\ref{pot}) corresponds to the limit of
only one light Higgs boson of the MSSM, and the values of the parameters
correspond to the light right-handed stop scenario.

\subsection{An upper bound}

The temperature $T_{N}$ at which nucleation of bubbles begins is defined as
that at which the first bubble is formed inside a causal volume \cite{ah92}.
In the radiation dominated era the horizon size scales like $d_{H}=2t$,
where the age of the Universe $t$ is given by Eq.~(\ref{ttrad}). Hence the
size of a causal volume is $V_{H}\sim 8\xi ^{3}M_{Pl}^{3}/T^{6}$, and the
probability of nucleating a bubble inside it is
\begin{equation}
P\left( T\right) =\int_{t_{c}}^{t}\Gamma V_{H}dt=\left( 2\xi M_{Pl}\right)
^{4}\int_{T_{c}}^{T}e^{-S_{3}/T}\frac{dT}{T^{5}}\ ,
\end{equation}
where we have used Eqs. (\ref{gamma}) and (\ref{ttrad}) in the last step.
The temperature $T_{N}$ is thus determined by the condition $P\left(
T_{N}\right) \sim 1$. This can be solved either by an analytical
approximation \cite{ah92} or numerically, and gives
\begin{equation}
S_{3}\left( T_{N}\right) /T_{N}\simeq 135\ .
\end{equation}
Although $T_{N}$ is often assumed to be the temperature of the transition,
it only corresponds to the onset of nucleation, so it gives an upper bound
for $T_{t}$.

Once bubbles begin to nucleate, however, it still takes them some time to
fill all space, since bubble walls cannot move faster than light. A less
conservative bound $T_{\max }<T_{N}$ can thus be obtained by setting  
$v_{w}=1 $ instead of using Eq.(\ref{veloc}) for $v_w$ in Eq.~(\ref{frac}). 
In this limit the expansion of bubbles is so fast that we can neglect their 
initial radius $r_{0}$. Then the condition that the fraction $f\left( 
T\right) $ falls to zero can be expressed as \begin{equation} \frac{4\pi 
}{3}\left( 2\xi M_{Pl}\right) ^{4}\int_{T}^{T_{c}}e^{-S_{3}\left( T^{\prime 
}\right) /T^{\prime }}\left( T^{\prime }-T\right) ^{3}\frac{ dT^{\prime 
}}{T^{\prime 5}}\sim 1\ , \end{equation} giving for the upper bound $T_{\max 
}$ \begin{equation} S_{3}\left( T_{\max }\right) /T_{\max }\simeq 105-109.  
\end{equation} The small uncertainty is due to the variation of the 
parameters of the potential as we take SM-like or MSSM-like values and the 
ratio $\lambda /E$ in the range $2-3$.

\subsection{A lower bound}

It is easy to see from Eqs. (\ref{s3}) and (\ref{f}) that the instanton
action $S_{3}$ vanishes at $T_{0}$. Hence, there is no exponential
suppression to the nucleation rate $\Gamma $ at that temperature (this is a
consequence of the disappearance of the barrier of $V\left( \phi \right) $
at $T_{0}$). At temperatures near $T_{0}$ the nucleation rate is thus $
\Gamma \sim T^{4}$, which is extremely large. Indeed, $\Gamma =T^{4}$
means that in a time interval $\tau \sim T^{-1}$ a bubble forms in every
volume $\mathcal{V}\sim T^{-3}$. This time $\tau $ is many orders of
magnitude less than the time elapsed between $T_{c}$ and $T_{0}$, Eq.~(\ref
{dtc0}). In addition, the size of the nucleated bubbles is larger than the
volume per bubble $\mathcal{V}$, since the radius of the electroweak bubble
is always larger than a few $T^{-1}$. So all space must be filled with
bubbles before the Universe cools down to some temperature $T_{\min }>T_{0}$
, when the exponential factor in Eq.~(\ref{gamma}) is still much less than $1
$.

We can calculate the lower bound $T_{\min }$ by setting $v_{w}=0$ in
Eq.~(\ref{frac}),
\begin{equation}
\frac{8\pi \xi M_{Pl}}{3}\int_{T_{\min }}^{T_{c}}\Gamma \left( T\right)
r_{0}^{3}\left( T\right) \frac{dT}{T^{3}}\sim 1\ .  \label{eqtmin}
\end{equation}
This corresponds to the (unrealistic) limit of a nucleation-dominated
first-order phase transition. The opposite limit is that of a bubble
expansion dominated phase transition, in which the nucleated bubbles grow so
rapidly that the original size is negligible, and the transition is
completed when the nucleation rate is still small \cite{ah92}. This is the
case of the upper bound $T_{\max }$ we have just estimated. As we shall see,
the electroweak phase transition is closer to an expansion dominated
transition than to a nucleation dominated one. The temperature $T_{\min }$
determined by Eq.~(\ref{eqtmin}) is given by $S_{3}\left( T_{\min }\right)
/T_{\min }\simeq 35-38$.

In terms of the relative variable $\varepsilon $, the bounds on the
temperature of the transition are expressed as $0.31\lesssim \varepsilon
_{t}\lesssim 0.52$ for SM values of $E$ and $D$, and $0.15\lesssim
\varepsilon _{t}\lesssim 0.31$ for MSSM values of the parameters (setting
for definiteness $\lambda =2E$ in both cases).

\section{Numerical results \label{develop}}

In the last section we have determined the temperatures $T_{\max }$ and $
T_{\min }$ between which the phase transition takes place. Depending on the
value of the friction parameter $\eta $, the temperature of the phase
transition will be closer to one of these bounds.

\subsection{The development of the phase transition}

\begin{figure}[tbh]
\centering
\epsfysize=5cm \leavevmode
\epsfbox{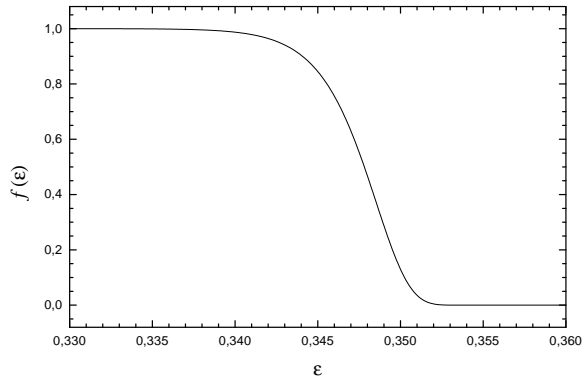}
\caption{Fraction of space in false vacuum as a function of $\varepsilon $,
for $\eta =100$, $D=0.2$, $E=0.006$ and $\lambda =2E$.}
\label{figfrac}
\end{figure}

The fraction of space that is still in the symmetric phase is obtained by
inserting Eq.~(\ref{veloc}) into Eq.~(\ref{frac}) and solving numerically
for $f\left( T\right) $. We have plotted $f\left( \varepsilon \right) $ in
Fig.~\ref{figfrac} for a given choice of parameters, in order to illustrate
the evolution of the phase transition. Note that, once the fraction of
volume occupied by bubbles becomes appreciable, the phase transition
completes in a tiny temperature interval. That is why we speak of the
``temperature of the transition'' rather than of a ``temperature interval''.
No relevant quantity will change appreciably in this interval. This is due
to the large time-temperature relation, which governs the nucleation and
expansion of bubbles. The duration of the phase transition varies very
little with the friction. It is given by
\begin{equation}
\frac{\Delta T_{t}}{T_{c}-T_{0}}\simeq \Delta \varepsilon _{t}\simeq 5\times
10^{-3}\
\end{equation}
for values of $\eta $ as large as $10^{5}$ as well as for no friction at all.

\begin{figure}[tbh]
\centering
\epsfysize=5cm \leavevmode
\epsfbox{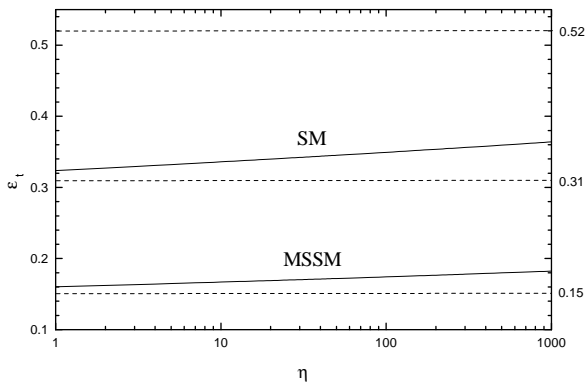}
\caption{The temperature of the transition as a function of the friction $
\eta $. The upper curve corresponds $D=0.2$, $E=6\times 10^{-3}$, and $
\lambda =2E$. The lower one is for $D=1$, $E=6\times 10^{-2}$. The
temperatures corresponding to upper and lower bounds are indicated with
dashed lines.}
\label{tteta}
\end{figure}

With no friction, the phase transition occurs at $T_{\max }$. When friction
is turned on, the transition is retarded, as shown in Fig.~\ref{tteta}. Note
that the dependence of $\varepsilon _{t}$ with $\eta $ is almost
logarithmic, so $\varepsilon _{t}$ increases more slowly for large $\eta $.
As a consequence, $T_{t}$ will be closer to $T_{\max }$ rather than to $
T_{\min }$ for not extremely high values of $\eta $. As discussed in section
\ref{growth}, $\eta $ should not be out of the range $1-1000$.

\begin{figure}[tbh]
\centering
\epsfysize=5cm \leavevmode
\epsfbox{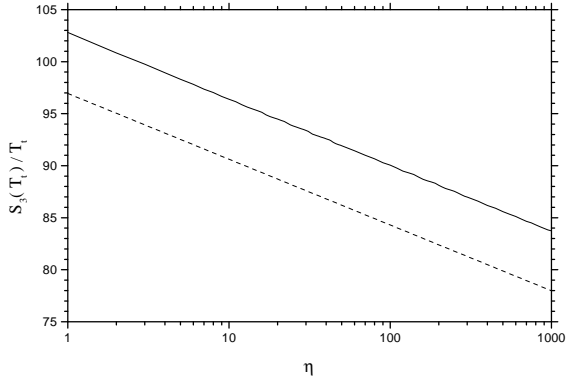}
\caption{The value of the three dimensional action $S_{3}$ at the
temperature of the transition, for $D=1$, $E=0.06$ and $\lambda =2E$. The
dashed line shows the maximum shift of the curve as $D$, $E$ and $\lambda $
are varied as indicated in Sec.~\ref{temp}. }
\label{s3teta}
\end{figure}

In Fig.~\ref{s3teta} we have plotted the value of $S_{3}/T$ at the moment of
the transition, as a function of the damping coefficient. The solid line
corresponds to MSSM values of the parameters. For likely values of the
friction, $\eta \sim 100$, the temperature of the transition is given by $
S_{3}\left( T_{t}\right) /T_{t}\simeq 90$. The dashed line indicates the
variation in $S_{3}/T$ when considering different parameters of the
effective potential, as explained in the previous section. Since the
ambiguity in $S_{3}\left( T_{t}\right) /T_{t}$ produced by such a variation
is of only a 5\%, we believe that Fig.~\ref{s3teta} not only describes
qualitatively the general case, but it would also give a good approximation
for the determination of the temperature of the electroweak phase transition
within different models.

\subsection{Implications for baryogenesis}

The amount of baryon number generated at the electroweak phase transition
depends on the shape of bubbles. We have seen that the relevant parameters,
namely, the width and velocity of the bubble wall and the value of the Higgs
field in the broken symmetry phase, vary significantly between the
temperatures $T_{c}$ and $T_{0}$, and hence the importance of determining
their values at the intermediate temperature $T_{t}$ at which the phase
transition occurs.

Regarding the wall velocity, we have shown in section \ref{growth} that,
although $v_{w}$ takes all the possible values $0<v_{w}<1$ in the interval $
0<\varepsilon <1$, it changes rapidly at the beginning and at the end of
that interval, but remains of the same order of magnitude for intermediate
values of $\varepsilon $, roughly between $\varepsilon =0.1$ and $
\varepsilon =0.8$ (see Fig.~\ref{vw}). Since the bounds on $T_{t}$ obtained
in section \ref{temp} fall in this region of slow variation of $v_{w}$, we
see that the wall velocity depends weakly on the actual temperature of the
transition, although it is sensitive to the friction, as shown in Fig.~\ref
{vw}. Therefore, rough approximations of the temperature of the transition,
such as the usual $T_{t}\simeq T_{N}$, do not introduce significant errors
in the estimation of $v_{w}$. The same can be said of the wall thickness, as
inferred from Fig.~\ref{figrl}.

Typical values of these quantities are found by taking for instance $\eta =70
$ and MSSM-like parameters for $\Delta V$. Then we obtain $l_{w}\left(
T_{t}\right) \simeq 16T^{-1}$, which inserted into Eq.~(\ref{veloc}) gives $
v_{w}\left( T_{t}\right) \simeq 3\times 10^{-3}$. This is in agreement with
the value obtained in Ref.~\cite{js00}, and supports an MSSM scenario for
electroweak baryogenesis.

With respect to the Higgs mean value $v\left( T\right) $, although it varies
by a 50\% between $T_{c}$ and $T_{0}$, the variation between the critical
temperature and $T_{t}$ will be smaller. As shown in Fig~\ref{tteta}, the
temperature of the transition stays close to the upper bound (lower bound on
$\varepsilon _{t}$), even for large values of $\eta $. We can use Eq.~(\ref
{palfa}) to estimate by which amount the order parameter can change.
Setting
for definiteness $\lambda =2E$, like in Fig.~\ref{tteta}, we see that for $
\eta \sim 100$ the upper curve in the figure gives $\varepsilon _{t}\simeq
0.35$. Since $v\left( T\right) /T=\left( E/\lambda \right) p\left( \alpha
\right) $ and $\alpha \simeq 1-\varepsilon $, we readily obtain $v\left(
T_{t}\right) /T_{t}\gtrsim 1.2$. Similarly, for the lower curve we obtain $
v\left( T_{t}\right) /T_{t}\gtrsim 1.1$. Comparing with $v\left(
T_{c}\right) /T_{c}=1$, we find a difference of at least a $10-20\%$
between the critical temperature and the actual temperature of the
transition. This means that the use of Eq.~(\ref{washout}) can result in an
upper bound on the Higgs mass that is by that percentage more severe than 
the correct constraint. A more precise bound for $\lambda $ can be obtained 
by using Eq.~(\ref{lambdae}) recursively. For instance, after computing $
\varepsilon _{t}$ with the assumption $\lambda =2E$
and calculating the constraint $
\lambda =p\left( \varepsilon _{t}\right) E$,
this new value of $\lambda$  can be used to recalculate $
\varepsilon _{t}$ and obtain a higher order correction to the bound.
This may be useful when considering a more concrete model for the phase
transition.

The error in the determination of $v\left( T_{t}\right) /T_{t}$ diminishes
considerably when using the temperature $T_{N}$ of the onset of nucleation
instead of $T_{c}$. For the values of the parameters that we have
considered, $v\left( T\right) /T$ increases by less than a 5\% between $T_{N}
$ and $T_{t}$. Hence, regarding the constraints that result by avoiding the
washout of the baryon asymmetry, we again find that $T_{t}\simeq T_{N}$ is a
good approximation in the calculations of electroweak baryogenesis.

Before concluding, we would like to discuss the behavior of another bubble
parameter that is relevant for baryogenesis in the MSSM, that is, the
relative variation of the two Higgs fields across the wall, $\Delta \beta $.
The $CP$-violating force on the quarks depends on this variation, and thus
the final baryon asymmetry results proportional to $\Delta \beta $. In order
to investigate the temperature dependence of this parameter, however, we
would need to consider the full two-Higgs model. The computation of the
bubble profile with more than one scalar field is a difficult task (see for
instance \cite{mqs98,j99}). Nevertheless, our results for the temperature of
the transition can be used to verify the approximations employed in existing
calculations of $\Delta \beta $. Moreno \emph{et.al.} \cite{mqs98}, for
example, have calculated the Higgs profile of the bubble
in the light stop scenario. They performed the calculation
at the temperature we have called $T_{N}$, and also at $
T_{N}+0.4GeV$ and $T_{N}-0.4GeV$, and the variation they obtained of
$\Delta \beta $ along the bubble wall
is very similar among these temperatures.
Translating that temperature interval to the
variable $\varepsilon $ by $\Delta \varepsilon \simeq \Delta T/\left(
T_{c}-T_{0}\right) $, where $\Delta T=0.8GeV$ and $T_{c}-T_{0}\simeq 3GeV$,
gives $\Delta \varepsilon \simeq 0.2$. So the conclusion is that the
parameter $\Delta \beta $ does not change significantly in an ample interval
around $\varepsilon _{N}$, which, according to our results,
includes the possible values of $ \varepsilon _{t}$.

\section{Conclusions and discussion \label{discus}}

It is well know that phase transition dynamics plays an important role in
electroweak baryogenesis. Although the production of the baryonic asymmetry
of the Universe at the electroweak phase transition is an exciting
possibility, it is clear that a complete picture of the phase transition is
needed to check quantitatively this possibility. This includes the detailed
investigation of the evolution of bubbles, as well as the accurate
calculation of the temperature of the transition. In this paper we have
investigated both of this problems, which are certainly related with each
other.

We have determined the upper and lower temperature limits $T_{\max }$ and $
T_{\min }$ inside the interval $T_{0}-T_{c}$, between which the phase
transition must take place regardless of the velocity of expansion of
bubbles. Our main conclusion in this respect is that, for the electroweak
phase transition, the lower bound on $T_{t}$ avoids the possibility that the
phase transition occurs at a temperature near $T_{0}$. This is good for
baryogenesis, since in such a case the transition would proceed by an almost
homogeneous growing of the Higgs field, what would lead to the generation of
a smaller baryon abundance.

When the evolution of the bubble wall is included, the temperature of the
phase transition varies between the limits $T_{\max }$ and $T_{\min }$ as
the friction parameter $\eta $ goes from $0$ to $\infty $. However, it turns
out that, due to the slow decreasing of temperature with time, the
temperature $T_{t}$ stays close to $T_{\max }$ for not extremely high values
of $\eta $. According to our numerical computations, for likely values of
the viscosity of the plasma the temperature of the electroweak phase
transition is given by $S_{3}\left( T_{t}\right) /T_{t}\lesssim 90$.

On the other hand, the values of the wall thickness $l_{w}\left( T\right) $
and wall velocity $v_{w}\left( T\right) $ are rather insensitive to
temperature in the range $T_{\min }-T_{\max }$, so they depend weakly on the
actual temperature of the phase transition. Regarding the Higgs mean value $
v\left( T\right) $, although baryogenesis is very sensitive to this
parameter, and hence the uncertainty between $T_{\min }$ and $T_{\max }$
would still be considerable, the result that the phase transition takes
place near $T_{\max }$ eliminates this uncertainty. Furthermore, the
temperature $T_{\max }$ is in general close to that of the onset of
nucleation. Hence we conclude that $l_{w}\left( T_{t}\right) $, $v_{w}\left(
T_{t}\right) $, and $v\left( T_{t}\right) $, which are relevant parameters
for baryogenesis, do not change significantly from those obtained with the
usual estimation of the temperature of the transition, i.e., with the
condition $S_{3}\left( T\right) /T\sim 135$. We have also shown that the
same applies to other quantities, which we have not computed, such as the
parameter $\Delta \beta $, which is relevant for electroweak baryogenesis in
the context of the MSSM.

Regarding further development of this analysis, a detailed investigation of
the effect of reheating due to the latent heat released by the expanding
bubbles would be necessary \cite{reheat}. As we have discussed in section
\ref{temp}, such an investigation would be especially important in the case
of the MSSM, where the main consequence of this effect would be to prevent
the wall velocity to reach the value $\sim 10^{-3}-10^{-2}$ that is needed
for a satisfactory result of electroweak baryogenesis.

\section*{Acknowledgments}

I would like to thank Prof. Luis Masperi for helpful comments. Part of
this work
was done with the support of a fellowship of the Consejo Nacional de
Investigaciones Cient\'{\i}ficas y T\'ecnicas, Argentina.

\appendix

\section{The shape of bubbles}

In this appendix we give some interesting details of the calculation of the
bubble radius, the wall width, and the value of the Higgs field inside the
bubble at the moment of formation.

The shape of the bubble is given by Eq.~(\ref{ecfi}) with potential (\ref
{pot}),
\begin{equation}
\frac{d^{2}\phi }{dr^{2}}+\frac{2}{r}\frac{d\phi }{dr}=2D\left(
T^{2}-T_{0}^{2}\right) \phi -3ET\phi ^{2}+\lambda \phi ^{3}\ .
\label{ecfiap}
\end{equation}
Following Dine et al. \cite{dlhll92}, we define $\phi =\left[ D\left(
T^{2}-T_{0}^{2}\right) /\left( ET\right) \right] \Phi $ and $r=R/\sqrt{
2D\left( T^{2}-T_{0}^{2}\right) }$, so that Eq.~(\ref{ecfiap}) becomes
\begin{equation}
\frac{d^{2}\Phi }{dR^{2}}+\frac{2}{R}\frac{d\Phi }{dR}=\Phi -\frac{3}{2}\Phi
^{2}+\frac{1}{2}\alpha \Phi ^{3}\ ,  \label{ecfiap2}
\end{equation}
with $\alpha =\lambda D\left( T^{2}-T_{0}^{2}\right) /\left( ET\right)
^{2}$. We have solved numerically this equation with the usual
overshooting-undershooting method, to obtain the configuration $\Phi \left(
R\right) $ for different values of $\alpha $. The graphics of Fig.~\ref
{shape} for $\phi /T$ as a function of the dimensionless radius $Tr$ are
then given by
\begin{equation}
\frac{\phi \left( T\right) }{T}=\frac{E\alpha }{\lambda }\Phi \left( \sqrt{
\frac{2E^{2}\alpha }{\lambda }}Tr\right) \ .
\end{equation}
The expression (\ref{s3}) for $S_{3}\left( T\right) /T$ is then obtained by
an appropriate integration of the solution over $d^{3}X=4\pi R^{2}dR$.

We define the radius $R_{0}\left( \alpha \right) $ of the configuration as
that at which $\Phi \left( R\right) $ falls to $\Phi _{0}/2$, where $\Phi
_{0}\left( \alpha \right) =\Phi \left( R=0\right) $. Similarly, we calculate
the radii $R_{1}$ and $R_{2}$ corresponding to $0.9\Phi _{0}$ and $0.1\Phi
_{0}$ respectively, and define the wall width as $L_{w}=R_{2}-R_{1}$. This
quantities are related to the physical radius $r_{0}$ and width $l_{w}$ of
Fig.~\ref{figrl} by

\begin{equation}
Tr_{0}=\frac{\sqrt{\lambda /2}}{E}\frac{1}{\sqrt{\alpha }}R_{0}\left( \alpha
\right) \ ,\quad Tl_{w}=\frac{\sqrt{\lambda /2}}{E}\frac{1}{\sqrt{\alpha }}
L_{w}\left( \alpha \right) \ .  \label{rl}
\end{equation}

The value of the field inside the bubble, plotted in Fig.~\ref{figfi}, is
defined as $\phi _{0}=\phi \left( r=0\right) $ and is related to the
dimensionless $\Phi _{0}$ by
\begin{equation}
\frac{\phi _{0}\left( T\right) }{T}=\frac{E}{\lambda }\alpha \Phi _{0}\left(
\alpha \right) \ ,
\end{equation}
to be compared with the minimum $v\left( T\right) $ of the potential, given
by
\begin{equation}
\frac{v\left( T\right) }{T}=\frac{E}{\lambda }p\left( \alpha \right) \ .
\end{equation}

The divergence of $r_{0}$ and $l_{w}$ at $\alpha =0$ is apparent in Eq.~(\ref
{rl}). The divergence of $r_{0}$ at $\alpha =1$, instead, is hidden in the
function $R_{0}\left( \alpha \right) $. The wall width is finite at $\alpha
=1$, but due to the divergence of the radius the shape of the bubble wall
cannot be calculated using the numerical method. Interestingly, in this
limit $L_{w}$ can be obtained analytically, what is useful to check that the
numerical curve goes to the right limit for $\alpha \rightarrow 1$: Since $
R_{0}\rightarrow \infty $ the second term disappears from
Eq.~(\ref{ecfiap2}). Then multiplying by $\frac{d\Phi }{dR}$ and
integrating over $dR$ gives for $\alpha =1$ \begin{equation}
\frac{d\Phi }{dR}=-\Phi \left( 1-\frac{\Phi }{2}\right) \end{equation}
Integrating again across the wall we obtain \begin{equation}
L_{w}=\left. \ln \left( \frac{\Phi -2}{\Phi }\right) \right| _{0.9\Phi
_{0}}^{0.1\Phi _{0}}\ .  \label{aplw}
\end{equation}
Similarly we can see that $\Phi \left( R\right) $ goes from $2$ to $0$
across the wall, so in this case $\Phi _{0}=2$, and Eq.~(\ref{aplw}) gives $
L_{w}=2\ln 9$. We can then verify that in Fig.~\ref{figrl}, with $\lambda =2E
$ and $E=0.06$, the wall width has the correct value at $\varepsilon =0$.

\end{document}